\documentclass[prb,twocolumn, amsmath,amssymb]{revtex4}
\setcitestyle{numbers,square}
\usepackage{amsmath,amssymb}
\usepackage{tabularx}
\usepackage{graphicx}
\usepackage{bmpsize}
\usepackage{bm}
\usepackage{color}
\usepackage{amssymb}
\usepackage[version=3]{mhchem}
\usepackage{newtxmath}
\DeclareMathAlphabet{\mathpzc}{OT1}{pzc}{m}{it}

\begin{document}

\title{Perfect one-dimensional interface states in a twisted stack of three-dimensional topological insulators}
\author{Manato Fujimoto}
\affiliation{Department of Physics, Osaka University,  Osaka 560-0043, Japan}
\author{Takuto Kawakami}
\affiliation{Department of Physics, Osaka University,  Osaka 560-0043, Japan}
\author{Mikito Koshino}
\affiliation{Department of Physics, Osaka University,  Osaka 560-0043, Japan}
\date{\today}

\begin{abstract} 
We theoretically study the electronic structure of interface states in twisted stacks of three-dimensional topological insulators. 
When the center of the surface Dirac cone is located at a midpoint of a side of BZ boundary, we find that an array of nearly-independent one-dimensional channels is formed by the interface hybridization of the surface states,
even when the moir\'{e} pattern itself is isotropic.
The two counter-propagating channels have opposite spin polarization, and they are robust against scattering by spin-independent impurities. 
The coupling between the parallel channels can be tuned by the twist angle.
The unique 1D states can be understood as effective Landau levels where the twist angle works as a fictitious magnetic field. 
\end{abstract}

\maketitle
\section{Introduction}

Moir\'e superlattices, formed by stacking two dimensional (2D) materials with a small twist angle or lattice mismatch, have become a topic of great interest in condensed matter physics.
In these materials, 
\cite{li2015observation,yeh2016direct, tong2017topological,naik2018ultraflatbands,lee2019theory,liu2020tunable,shen2020correlated,lian2020flat,weston2020atomic,rosenberger2020twist,wang2020correlated,kennes2020one,fujimoto2021effective,an2021emergence,luo2021magic,he2021symmetry,park2021tunable,song2021direct,akram2021moire,xu2021emergence,xie2022twist,chaudhary2022twisted,wang2022one,PhysRevB.105.165422}
highly tunable electronic structures by twist angle
serve as platforms to search for novel exotic phenomena.
In a low-angle twisted bilayer graphene, in paticular, Dirac fermions of intrinsic graphenes are strongly hybridized into nearly-flat bands \cite{PhysRevLett.99.256802,PhysRevB.81.161405,trambly2010localization,PhysRevB.81.165105,PhysRevB.82.121407,bistritzer2011moire,PhysRevB.83.045425,PhysRevB.84.075425,dos2012continuum,PhysRevB.85.195458,PhysRevB.86.125413,PhysRevB.87.205404}, 
leading to strongly correlated phenomena such as superconductivity and strongly correlated insulating states.
Moir\'e superlattices have been studied in a wide variety of 2D materials, including twisted multilayer graphenes\cite{lee2019theory,liu2020tunable,shen2020correlated,he2021symmetry,park2021tunable}, transition metal dichalcogenides \cite{yeh2016direct, naik2018ultraflatbands,weston2020atomic,rosenberger2020twist,wang2020correlated,wang2022one}, 2D magnets \cite{song2021direct,akram2021moire,xu2021emergence,xie2022twist} and semiconductors \cite{li2015observation,kennes2020one,fujimoto2021effective,an2021emergence}.

The exploration of moir\'e physics was also extended to three-dimensional topological insulators (3DTIs), which contain two-dimensional Dirac fermions on the surface. 
For instance, the moir\'e pattern was observed on misaligned topmost quintuple layers of Bi$_2$Se$_3$\cite{liu2014tuning} and Bi$_2$Te$_3$\cite{schouteden2016moire}, and the electronic property in such a system was theoretically studied.\cite{PhysRevX.11.021024}
Previous theoretical works also investigated a 3DTI modulated by 2D materials placed on the surface \cite{PhysRevB.103.155157}, and topological phases in a twisted bilayer of Bi$_2$(Te$_{1-x}$Se$_x$)$_3$.\cite{tateishi2021quantum}

Here we ask: when a pair of 3DTI slabs are overlapped on top of each other with a twist angle, how the surface Dirac cones would interact. 
One may think that physics would be similar to the twisted bilayer graphene, which also have Dirac cones with moir\'e interface coupling.
In this paper, on contrary to the expectation, we demonstrate that the interface band structures are completely different depending on the position of the Dirac cone in the Brillouin zone (BZ).
Generally, the center of the surface Dirac cone is located at a time reversal invariant momentum (TRIM).
When the TRIM is at $\Gamma$ point, the surface states are found to be simply gapped out in a twisted interface.
When the TRIM is at a corner point of BZ, 
a flat band similar to twisted bilayer graphene is realized.\cite{dunbrack2021magic}

\begin{figure}
  \begin{center}
 \includegraphics[width=1.0 \hsize]{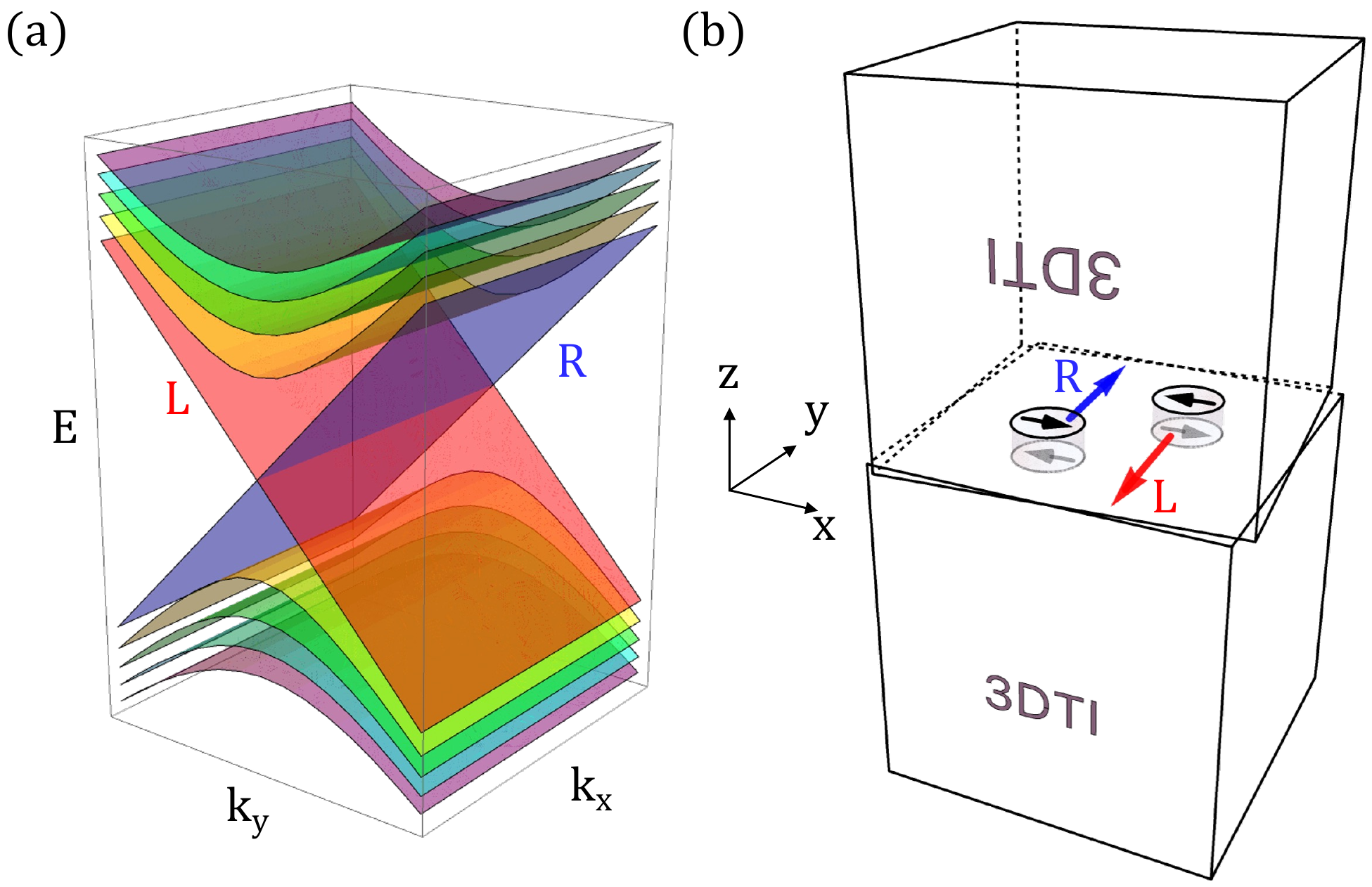}
    \caption{
    (a)Band structure of the interface states in twisted 3DTIs with the TRIM at a midpoint of the Brillouin zone boundary. The energy bands are completely flat in $k_x$, while disperse in $k_y$ direction. ``L'' and ``R'' represent linear bands with left-going and right-going band velocities.
    (b) Schematic picture of L and R states where red and blue arrows indicate the corresponding band velocities, and black arrows represent spin directions in upper and lower 3DTI's surfaces.
}
    \label{fig_3Dband}
  \end{center}
  \end{figure}

When the TRIM is at a midpoint of a side of BZ boundary,
on the other hand, we find that the surface Dirac cones of top and bottom surfaces are converted to a perfect 1D channel as depicted in Fig.~\ref{fig_3Dband}(a), which is completely flat in one direction while disperses in the other direction. There, the left-going and right-going modes have opposite spin textures [Fig.~\ref{fig_3Dband}(b)], and they are robust against scattering by spin-independent impurities. 
Moreover, these peculiar 1D bands can be analytically understood as Landau levels of a formally equivalent lattice model, where the twist angle works as a fictitious magnetic field. The 1D bands are formed when the twist angle is sufficiently small, and there is no magic angle condition to have the 1D channels.
The theory is applicable to the side-centered surface Dirac cones found in SnTe $(111)$ surface. \cite{tanaka2012experimental, PhysRevB.90.235114, PhysRevB.89.125308} 
The interaction between 1D channels can be tuned by the twist angle, and it would be an ideal platform to study weakly-coupled Tomonaga-Luttinger liquid 
 \cite{wen1990metallic,emery2000quantum,sondhi2001sliding,vishwanath2001two,mukhopadhyay2001sliding, akram2021moire, kane2002fractional,teo2014luttinger,tam2021nondiagonal,neupert2014wire,iadecola2016wire,meng2015coupled,patel2016two}.


The paper is organized as follows. In Sec. \ref{sec_eff_3DTI}, we introduce a continuum model of twisted 3DTIs based on the symmetry consideration. 
In Sec. \ref{sec_M}, we calculate the energy spectrum for the M-centered model, and we describe the emergence of the spin polarized 1D propagation mode by a pseudo-Landau description.
A brief conclusion is given in Sec. \ref{sec_concl}.

\section{Effective continuum model}\label{sec_eff_3DTI}

 \begin{figure*}
  \begin{center}
 \includegraphics[width=1.0 \hsize]{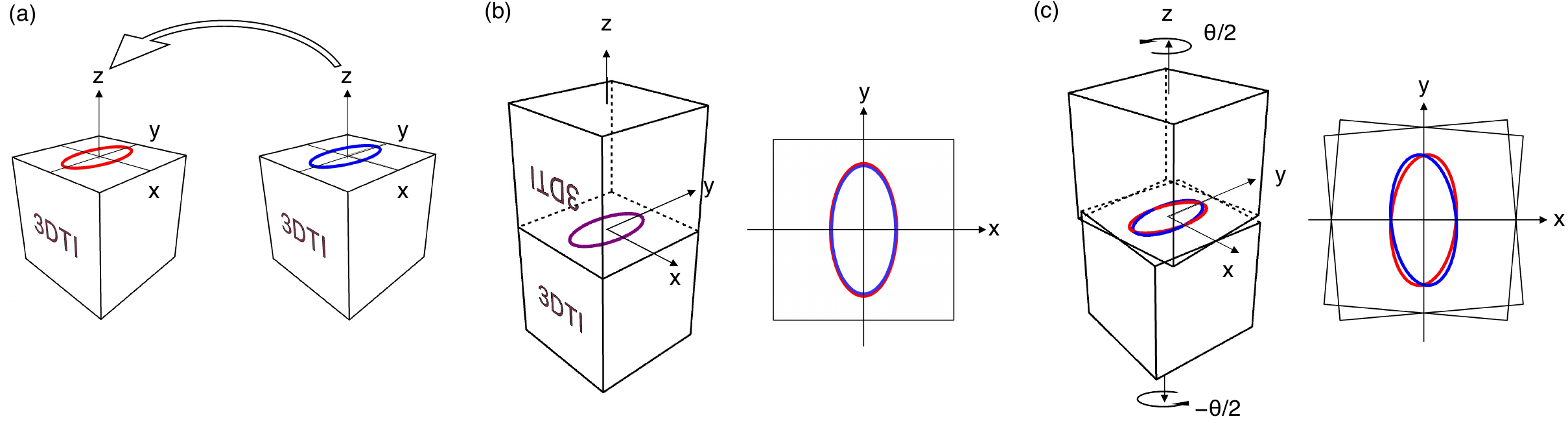}
    \caption{
    Construction of a twisted 3DTI stack.
(a) A pair of identical 3DTI slabs.
An ellipsoid on each surface represents the mirror symmetries $\mathcal{M}_{x}$ and $\mathcal{M}_{y}$.
(b) Non-twisted 3DTI stack. One slab is flipped over and stacked on the other such that the two slabs share the same mirror planes.
(c) Twisted 3DTI stack with the twist angle $\theta$.
}
    \label{fig_schem}
  \end{center}
  \end{figure*}

 \begin{figure}
  \begin{center}
 \includegraphics[width=1.0 \hsize]{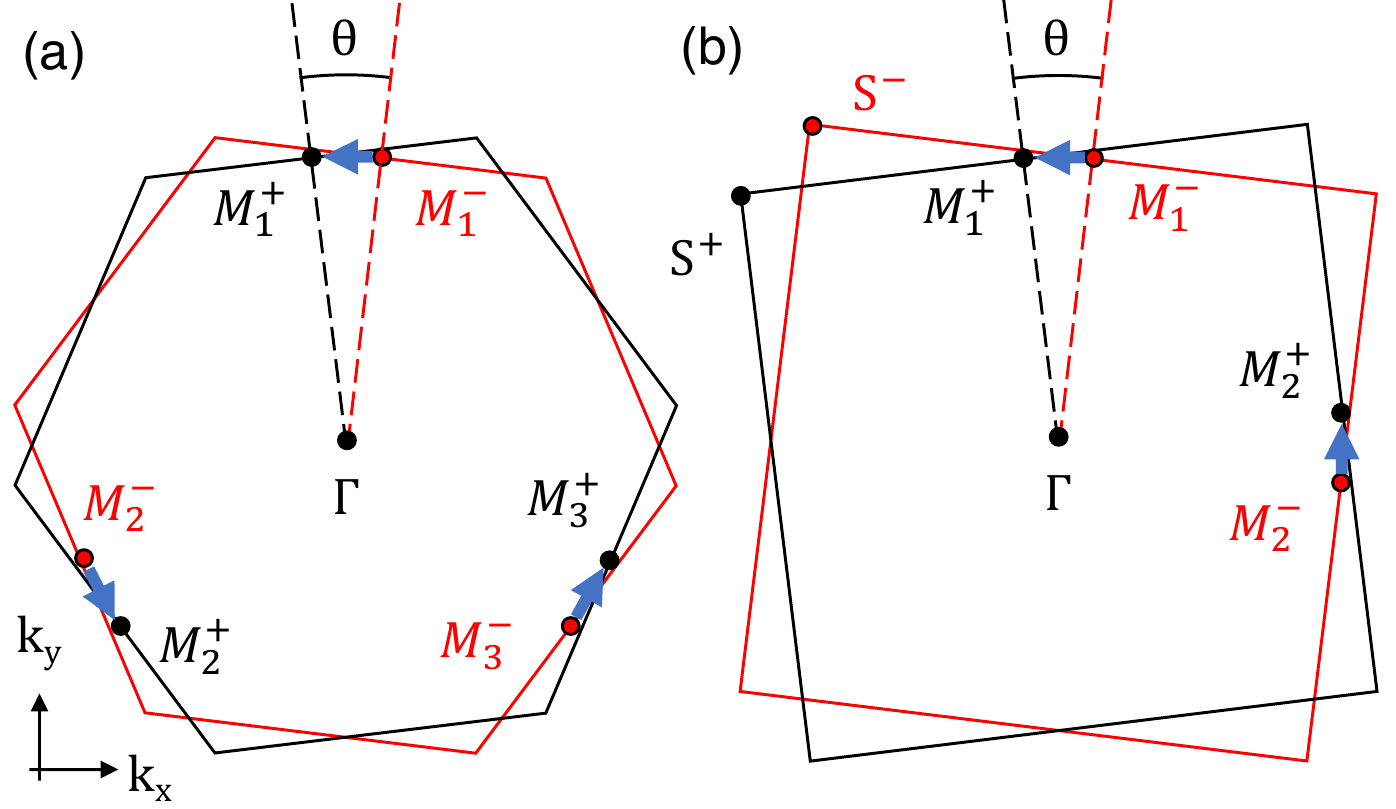}
    \caption{
The first BZs and TRIMs of (a) trigonal/hexagonal and (b) tetragonal crystals twisted by $\pm\theta/2$.
The TRIMs are shown as dots. 
}
    \label{fig_mBZ}
  \end{center}
  \end{figure}


We construct a twisted 3DTI stack in the following procedures.
We consider a pair of the same kind of 3DTI slabs shown in Fig.~\ref{fig_schem}(a).
We set $xy$ plane on the surface, and $z$ axis perpendicular to it.
We assume the individual 3DTI slabs (before twisting) have (1) time reversal symmetry $\mathcal{T}$ and (2) mirror symmetries of
$\mathcal{M}_{x}$ (with respect to $yz$-plane) and 
$\mathcal{M}_{y}$ ($zx$-plane).
We flip one slab and stack it on the top of the other,
such that the two slabs share the same mirror planes [Fig.~\ref{fig_schem}(b)].
Then we twist the top and bottom slabs with respect the $z$-axis 
by angle $\pm\theta/2$, respectively, and finally obtain a twisted 3DTI system [Fig.~\ref{fig_schem}(c)].
The entire system has time reversal symmetry and two-fold rotation symmetries along $x,y,z$ axes, $C_{2x},C_{2y},C_{2z}$.
In Appendix \ref{sec_T}, we also present the consideration of 3DTIs with lower spatial symmetries where individual slabs respect  only either $\mathcal{M}_{x}$ or $\mathcal{M}_{y}$,
where we will see that the basic properties are not changed.

At the interface, the surface states of the individual TIs are coupled by the moir\'{e} interlayer coupling.
We derive an effective continuum model to describe these hybrid surface modes.
The Hamiltonian is generally written as
\begin{equation}\label{eq_general_Ham}
H=
\left(
\begin{array}{cc}
H_u & H_{\rm int}^{\dagger} \\
H_{\rm int} & H_l \\
\end{array}
\right),
\end{equation}
where $H_{u/l}$ is the surface Dirac Hamiltonian of the upper and lower
TIs, respectively, and $H_{\rm int}$ describes the interface coupling.

We determine the form of $H_{u/l}$ and $H_{\rm int}$ from the symmetry consideration without specifying a microscopic model.
For the intra-surface part, $H_{u/l}$,
the center of the surface Dirac cone 
is generally located at time reversal invariant momentum (TRIM) $\bm{\Lambda}$ in the BZ.
In this paper, we consider two distinct cases where the TRIM is 
$\Gamma$ point ($\bm{\Lambda}=0$) and $M$ point (the midpoint of a side of the BZ boundary).
From the symmetry constraints ($\mathcal{T},\mathcal{M}_{x},\mathcal{M}_{y}$),
the surface Dirac Hamiltonian $H_{u/l}$ before the twist is written in the lowest order in $k$ as
\begin{equation}\label{eq_surface_Dirac}
H_{u/l}=\pm (v_x s_2 k_x - v_y s_1 k_y),
\end{equation}
where $\pm$ are for $u$ and $l$, respectively,
$v_{x/y}$ is the band velocity in $x/y$ direction, $\bm{k}=(k_x,k_y)$ is two-dimensional momentum, 
and $s_i\,(i=1,2,3)$ is the Pauli matrices for spin degree of freedom.
The derivation for Eq.~\eqref{eq_surface_Dirac} is given in Appendix \ref{sec_deriv_Dirac}.
In the following, we assume an isotropic surface Dirac cone $v_x=v_y (\equiv v)$ for simplicity, 
while anisotropy does not much affect qualitative results as shown in Appendix \ref{sec_anisotropic}.

In the twisted TI stack, the TRIM is shifted to $R(\pm\theta/2)\bm{\Lambda}$ where $R(\alpha)$ is the rotation matrix on the surface plane by angle $\alpha$.
The location of the TRIMs are as illustrated in Fig.~\ref{fig_mBZ}(a) and (b) for the trigonal/hexagonal and the tetragonal cases, respectively.
Note that $M_1, M_2,\cdots$ are distinct points.
While the midpoints on the BZ boundary in the tetragonal system are usually labelled by $X$ and $Y$, we use symbol $M$ 
regardless of the crystal symmetry through out the paper.
The intra-surface Hamiltonians $H_{u/l}$ of the twisted TI stack 
(relative to the shifted TRIMs) are given by \eqref{eq_surface_Dirac} even after rotation, as long as $v_x=v_y$ is assumed.



 \begin{figure*}
  \begin{center}
    \includegraphics[width=0.85 \hsize]{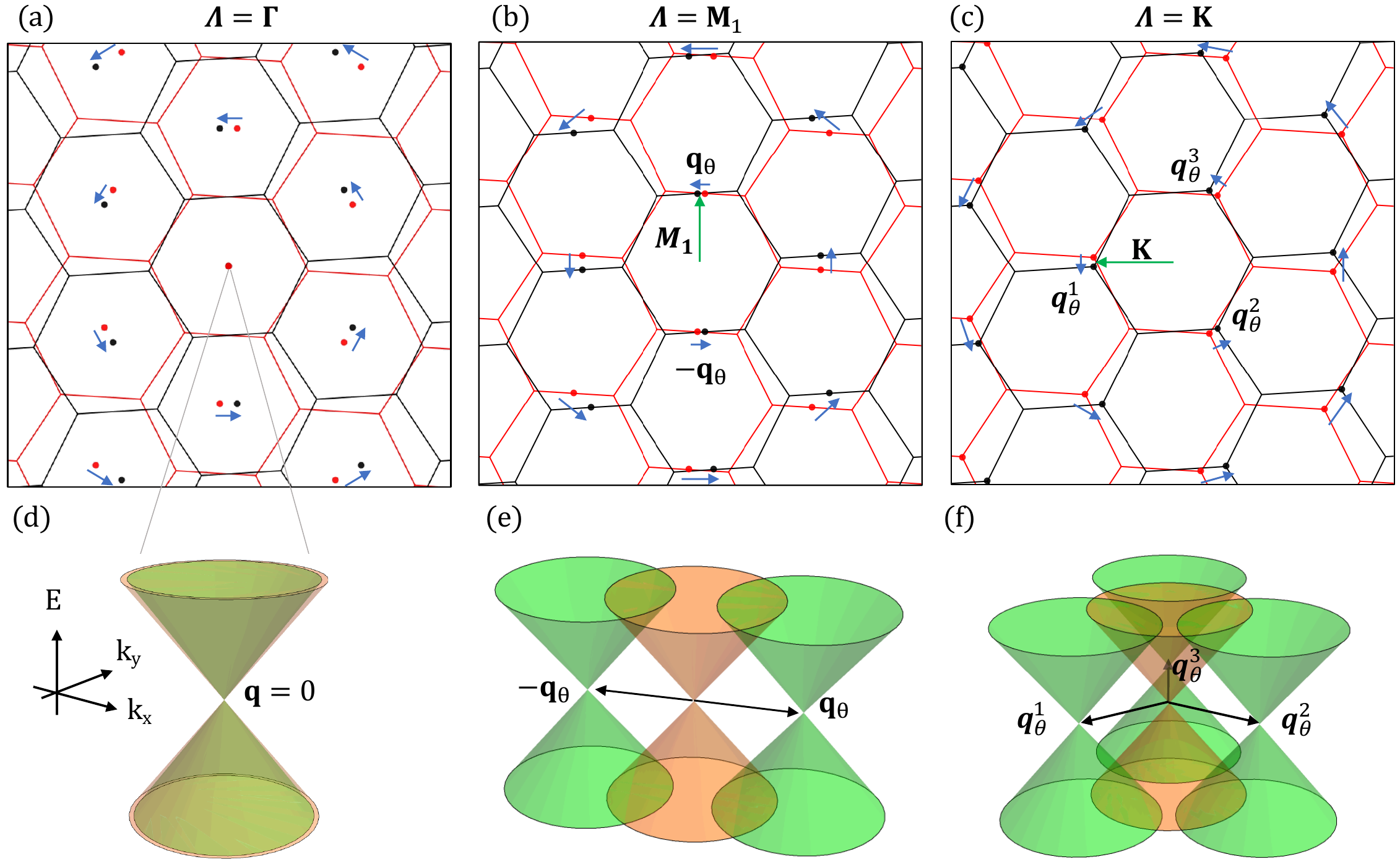}
    \caption{
    Equivalent points of (a)$\Gamma$, (b)$M_1$, (c)$K$ 
    in the twisted hexagonal BZs in the extended zone scheme.
    The red and black dots represent 
    $\bm{\Lambda}^\pm_{n_1,n_2}$ [Eq.~\eqref{eq_Lambda_pm}],
    and the blue arrows are the momentum shift vector $\bm{q}_{n_1,n_2}$ [Eq.~\eqref{eq_q_n1n2}].
    Lower panels (d)-(f) illustrate the coupling of the upper and lower Dirac cones with the dominant momentum shifts in (a)-(c), respectively.
}
    \label{fig_q_vec}
  \end{center}
  \end{figure*}

The interface-coupling Hamiltonian $H_{\rm int}$ can be obtained from the following consideration.
Let $\bm{b}_i\,(i=1,2)$ the reciprocal lattice vectors of the 3DTI without rotation. In the extended zone scheme, the equivalent points of $\bm{\Lambda}$ are written as
\begin{equation}
\bm{\Lambda}_{n_1,n_2} = \bm{\Lambda}+n_1\bm{b}_1+n_2\bm{b}_2,    
\end{equation}
where $n_1$ and $n_2$ are integers.
In the twisted TI stack, they are shifted to
\begin{equation}\label{eq_Lambda_pm}
    \bm{\Lambda}^\pm_{n_1,n_2} 
    = R(\pm\theta/2)\bm{\Lambda}_{n_1,n_2}.
\end{equation}
Figure \ref{fig_q_vec} shows the points of $\bm{\Lambda}^\pm_{n_1,n_2} $ for $\Gamma$, $M_1$ and $K$ point in the hexagonal lattice,
where red and black points represent $\pm$, respectively.
Note that $K$ is not a TRIM, but it is relevant to consider the interlayer coupling of twisted bilayer graphene 
and here we include it for comparison.
At each $(n_1,n_2)$, we can define an interlayer shift as
\begin{equation}\label{eq_q_n1n2}
    \bm{q}_{n_1,n_2}   =  \bm{\Lambda}^+_{n_1,n_2} -\bm{\Lambda}^-_{n_1,n_2} . 
\end{equation}

When $\theta$ is small, the interface Hamiltonian is approximately given by \cite{koshino2015interlayer}
\begin{equation}\label{eq_general_T}
H_{\rm int}=\sum_{n_1,n_2} T(\bm{\Lambda}_{n_1,n_2}) \exp(i\bm{q}_{n_1,n_2}\cdot\bm{r}),
\end{equation}
where $T(\bm{k})$ is a $2\times 2$ matrix.
In the tight-binding approach, $T(\bm{k})$ is essentially
the Fourier transform of the distance-dependent
interlayer transfer integral in the real space,
and therefore its amplitude rapidly decays in increasing $|\bm{k}|$.\cite{koshino2015interlayer}
The interface coupling Eq.~\eqref{eq_general_T} can be intuitively understood by using the diagram of Fig.~\ref{fig_q_vec}.
For each pair of black and red dots ($\bm{\Lambda}^+_{n_1,n_2},\bm{\Lambda}^-_{n_1,n_2}$),
the relative position $\bm{q}_{n_1,n_2}$
represents the shift of the Dirac-cone centers, and the distance from the origin 
(i.e., $\bm{\Lambda}_{n_1,n_2}$)
determines the coupling amplitude, where
a further distance corresponds to a smaller coupling.

In this paper, we only take the nearest black-red pair(s)
within the first BZ, which gives a dominant contribution.
For $\Gamma$ point  [$\bm{\Lambda}=0$, Fig.~\ref{fig_q_vec}(a)], for instance,
we only take the central pair with a shift of $\bm{q} = 0$.
For $M_1$ point [Fig.~\ref{fig_q_vec}(b)], we have two pairs at same distance from the origin,
which have the shift vectors of $\bm{q} =\pm \bm{q}_\theta$, where $\bm{q}_\theta = [R(\theta/2)-R(-\theta/2)]\bm{M}_1$.
For $K$ point [Fig.~\ref{fig_q_vec}(c)], there are the three nearest pairs with shifts of $\bm{q}=\bm{q}_\theta^1,\bm{q}_\theta^2,\bm{q}_\theta^3$, where
$\bm{q}_\theta^j= R[2\pi (j-1)/3]\Delta \bm{K}$
and $\Delta \bm{K} = [R(\theta/2)-R(-\theta/2)]\bm{K}$.
The coupling of the upper and lower surface Dirac cones with the shift vectors are illustrated in the lower panels of Fig.~\ref{fig_q_vec}.
The $\Gamma$-point case [Fig.~\ref{fig_q_vec}(d)] only has an interlayer coupling with $\bm{q} = 0$, i.e. no spatial modulation.
The $M$-point case [Fig.~\ref{fig_q_vec}(e)] only gives a one-dimensional coupling with $\bm{q} =\pm \bm{q}_\theta$.
The $K$-point case [Fig.~\ref{fig_q_vec}(f)] corresponds to the twisted bilayer graphene\cite{bistritzer2011moire,PhysRevB.83.045425,PhysRevB.84.075425,dos2012continuum,PhysRevB.85.195458,PhysRevB.86.125413,PhysRevB.87.205404,koshino2015interlayer}, where three shift vectors $\bm{q}_\theta^1,\bm{q}_\theta^2,\bm{q}_\theta^3$
give a two-dimensional superlattice coupling of the Dirac cones.\cite{bistritzer2011moire,PhysRevB.83.045425,PhysRevB.84.075425,dos2012continuum,PhysRevB.85.195458,PhysRevB.86.125413,PhysRevB.87.205404,koshino2015interlayer}

The specific form of $H_{\rm int}$ can be obtained by the symmetry consideration as follows.
First let us consider the $\Gamma$-point centered case, where the interface-coupling Hamiltonian is given by a constant matrix $H^{(\Gamma)}_{\rm int}=T$ (independent of the position $\bm{r}$) as argued.
For the Hamiltonian of twisted TI, Eq.~\eqref{eq_general_Ham}, the symmetry operators are expressed as $\mathcal{T}=i s_2 \mathcal{K} \otimes \tau_0$, $\mathcal{C}_{2x}= is_1 \otimes \tau_1$ and $\mathcal{C}_{2y}= is_2 \otimes \tau_1$, 
where $\mathcal{K}$ is the complex conjugate operator and $\tau_i$ is the Pauli matrix for $u/l$ degree of freedom.
We require that Eq.~\eqref{eq_general_Ham} respects
these symmetries, or
\begin{align}
&\mathcal{T}H(\bm{k},\bm{r})\mathcal{T}^{-1} = H(-\bm{k},\bm{r}),
\nonumber\\
&\mathcal{C}_{2x}H(\bm{k},\bm{r})\mathcal{C}_{2x}^{-1} = H(\mathcal{C}_{2x}\bm{k},\mathcal{C}_{2x}\bm{r}),
\nonumber\\
&\mathcal{C}_{2y}H(\bm{k},\bm{r})\mathcal{C}_{2y}^{-1} = H(\mathcal{C}_{2y}\bm{k},\mathcal{C}_{2y}\bm{r}),
\end{align}
where $\bm{k}$ and $\bm{r}$ are 2D vectors on the interface plane.
Then the inter-suraface submatrix $H_{\rm int}$ is forced to have the form,
\begin{equation}
    H^{(\Gamma)}_{\rm int} = t_0 s_0 + i t_3 s_3,
\end{equation}
where $t_0$ and $t_3$ are real numbers.
The entire Hamiltonian is written as
\begin{equation}
  H^{(\Gamma)}=v(s_2 k_x - s_1 k_y) \otimes\tau_3 + t_0 s_0\otimes \tau_1 + t_3 s_3\otimes \tau_2,
\end{equation}
giving the eigenvalues
$E=\pm(v^2k^2 \pm 2t_3 v k +t_0^2+t_3^2)^{1/2}$.
Therefore, the Dirac cones obtain an energy gap of the magnitude of $(t_0^2+t_3^2)^{1/2}$. This is a trivial result naturally expected when 
the same kind of TIs are overlapped.

When the surface Dirac cone is located at $M_1$ as shown in Fig.~\ref{fig_mBZ}, on the other hand, the interface-coupling Hamiltonian takes the form of $H^{(M_1)}_{\rm int}=T(\bm{M}_1)e^{i\bm{q}_\theta\cdot\bm{r}} + T(-\bm{M}_1) e^{-i\bm{q}_\theta\cdot\bm{r}}$.
By requiring that Eq.~\eqref{eq_general_Ham} respects 
$\mathcal{T}, \mathcal{C}_{2x}, \mathcal{C}_{2y}$,
the interface Hamiltonian Eq.~\eqref{eq_general_T} is reduced to
\begin{align}\label{eq_M_T1}
& H^{(M_1)}_{\rm int} = T_+ e^{i\bm{q}_\theta \cdot \bm{r}} + T_- e^{-i\bm{q}_\theta \cdot \bm{r}}, \nonumber\\
& T_\pm = t_0 s_0 \pm t_2 s_2 +i t_3 s_3,  
\end{align} 
where $t_0, t_2, t_3$ are real numbers.
As we will see in the following section, this results in a completely different band structure with nearly perfect one-dimensional modes.

\section{Perfect 1D states in M-point 3DTI}
\label{sec_M}
\subsection{Band structure}
\label{sec_band}

\begin{figure*}[t]
 \begin{center}
  \includegraphics[width=1 \hsize]{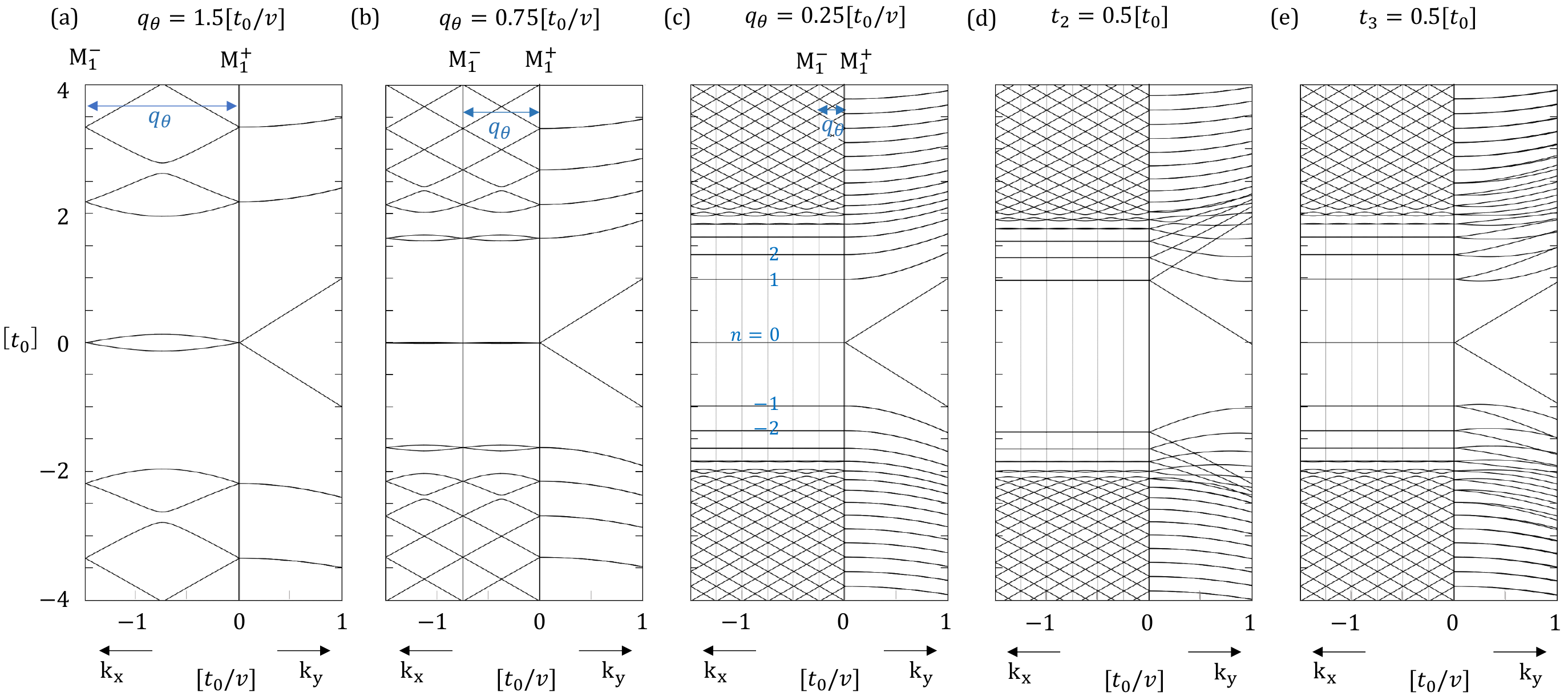}
  \caption{Band structure of the twisted 3DTIs with the surface Dirac cone at $M_1$,
  calculated for different twist angles of
   (a)$q_\theta=1.5$, (b)$0.75$, (c)$0.25$ (in units of $t_0/v$) with $t_2=t_3=0$,
  and for different band parameters
  (d)$(t_2,t_3)=(0.5t_0,0)$ and (e) $(0,0.5t_0)$ with $q_\theta=0.25$.
The left and right hand sides of each panel 
show the band dispersions along $k_x$ and $k_y$ axes, respectively, with respect to $M_1^+$.
For $k_x$, we take the extended zone scheme where the BZ boundaries are indicated by vertical lines.
 }
  \label{fig_band_C2xC2y}
 \end{center}
\end{figure*}

We calculate the band structure of the twisted 3DTIs with the surface Dirac cone at $M_1$.
Following the argument in the previous seciton, we consider the Hamiltonian, 
\begin{equation}\label{eq_M_Ham}
H=
\left(
\begin{array}{cc}
H_u(\bm{k})  & H^{(M_1)\dagger}_{\rm int} \\
H^{(M_1)}_{\rm int} &  H_l(\bm{k}) \\
\end{array}
\right),
\end{equation}
where $H_{u/l}(\bm{k}) = \pm v( s_2 k_x - s_1 k_y)$, and
$H^{(M_1)}_{\rm int}$ is given by \eqref{eq_M_T1}.
By arranging the wave bases as 
$\{\cdots,
|\bm{k}_{-1}\uparrow\rangle,|\bm{k}_{-1}\downarrow\rangle,
|\bm{k}_0\uparrow\rangle,|\bm{k}_0\downarrow\rangle,
|\bm{k}_1\uparrow\rangle,|\bm{k}_1\downarrow\rangle,
\cdots\}$ where $\bm{k}_n = \bm{k} + n \bm{q}_\theta$, 
the Hamiltonian matrix is written as
\begin{align}\label{eq_M_Ham_org}
&H=\nonumber\\
&
\begin{pmatrix}
\ddots \\
H_u(\bm{k}_{-2}) & T_{+}^{\dagger} &  \\
T_+  & H_l(\bm{k}_{-1}) & T_- \\
& T_{-}^{\dagger} & H_u(\bm{k}_0) & T_{+}^{\dagger} & \\
&& T_{+} & H_l(\bm{k}_{1}) & T_- \\
&&& T_{-}^{\dagger} & H_u(\bm{k}_2) \\
&&&& \ddots \\
\end{pmatrix},
\end{align}
with $T_\pm$ defined in Eq.~\eqref{eq_M_T1}.
This describes one-dimensional coupling in a series of the Dirac cones illustrated in Fig.~\ \ref{fig_q_vec}(e).
Here we set $\bm{q}_\theta = (q_\theta,0)$,
so that the moir\'e BZ is given by $-q_\theta/2 \leq k_x \leq q_\theta/2$,
while unbounded in $k_y$ direction.
The $q_\theta$ is related to the twist angle by 
\begin{equation}
q_\theta = 2M_1\sin\frac{\theta}{2} \approx M_1 \theta,
\label{eq_q_theta}
\end{equation}
where $M_1$ is the distance from $\Gamma$ to $M_1$.

We numerically calculate band structure by truncating the series with a sufficiently large $|n|$, and diagonalizing the Hamiltonian matrix.
We have three real parameters, $t_0, t_2, t_3$ in the interface Hamiltonian.
First we consider the simplest case with $t_2=t_3=0$.
The system is characterized by a dimensionless wavenumber $q_\theta/(t_0/v)$,
which is proportional to the twist angle $\theta$.
In Figs.~\ref{fig_band_C2xC2y} (a), (b) and (c),
we plot the band structures for $q_\theta = 1.5, 0.75, 0.25 (t_0/v)$ with $t_2=t_3=0$,
where the left and right hand sides of each panel 
show the band dispersions along $k_x$ and $k_y$ axes, respectively,
with the origin taken at $M_1^+$.
For $k_x$, we take the extended zone scheme where the BZ boundaries are indicated by vertical lines.

We see that the interface coupling opens a band gap at every crossing point of different Dirac cones,
whereas the energy bands at $M_{1}^\pm$ remain doubly degenerate, which are the Kramers doublets.
In decreasing $q_\theta$ (descreasing twist angle), the low-lying bands are flattened in $k_x$ axis, 
while the dispersion along $k_y$ remains intact.
The entire band structure on $k_xk_y$ plane is illustrated in Fig.~\ref{fig_3Dband}(a),
showing a nearly-perfect 1D bands which disperses only in the $y$ direction.
If we have inequivalent $M_j$ points as in Fig.~\ref{fig_mBZ},
each valley contributes to such a 1D band which disperses in parallel to the line $\Gamma-M_j$, while the bands of different $M_j$'s are not hybridized as long as the moir\'e period is much greater than the atomic scale.
Anisotropic band flattening in a moir\'e system with a base momentum at the BZ edge was also argued in a previous work from a crystallographic point of view.\cite{kariyado2019flat}

We label the 1D subbands as $n=0, \pm 1, \pm 2, \cdots$, as Fig.~\ref{fig_band_C2xC2y}(c).
In $k_y=0$, each level is doubly degenerate due to the 
Kramers doublet at $M_{1}^\pm$ and the band flattening.
The energy of the $\pm n$-th band in $k_y=0$ approximates $\pm\sqrt{4t_0 v q_\theta |n|}$,
which is analogous to the Landau levels of monolayer graphene.\cite{{miller2009observing,PhysRevLett.102.176804,song2010high}}
Actually there is a formal mapping of the Hamiltonian of twisted 3DTIs
[Eq.~\eqref{eq_M_Ham_org}] to a 2D Dirac Hamiltonian under a magnetic field,
which will be argued in the next section.

We plot the band strcture with nonzero $t_2$ and $t_3$  
in Fig.~\ref{fig_band_C2xC2y}(d) and (e), respectively,
at the same $q_\theta$ as in Fig.~\ref{fig_band_C2xC2y}(c).
The $t_2$ works like a mass in the graphene's Landau level shifting $n=0$ level by $2t_2$, and also splits the double degeneracy in $k_y\neq0$.
The $t_3$ hardly changes the band energies of $k_y=0$,
while it only splits the double degeneracy in $k_y\neq0$.
These features can also be explained by the mapping to the Landau levels.

\subsection{Pseudo-Landau level description for 1D interface modes}\label{sec_Landau}

To understand the 1D dispersion argued in the previous section,
we introduce a formal mapping between the $M$-centered twisted 3DTIs [Eq.~\eqref{eq_M_Ham_org}] and a 2D Dirac Hamiltonian under a magnetic field.
First, we apply a spin rotation to Eq.~\eqref{eq_M_Ham_org}
which replaces spin Pauli matrices as
\begin{equation}\label{eq_spin_rotation}
    (s_1, s_2, s_3) \to (s_2, s_3, s_1).
\end{equation}
As a result, the Hamiltonian matrix becomes,
\begin{widetext} 
\begin{equation}\label{eq_Ham1_sz_pm}
H=
\left(
\begin{array}{ccccc|ccccc}
                   \ddots &              &             &              &        &\ddots &              &             &           \\
                          & \epsilon_{-1}&   t_0-t_2   &              &        &       &     -ivk_y   &     it_3    &              &\\
                          &    t_0-t_2   & \epsilon_{0}&   t_0+t_2    &        &       &     -it_3    &     ivk_y   &     -it_3    &\\
                          &              &   t_0+t_2   & \epsilon_{+1}&        &       &              &     it_3    &     -ivk_y    &\\
                          &              &             &              & \ddots &       &              &             &              &\ddots\\
                                                                            \hline
                   \ddots &              &             &              &        &\ddots &              &             &              &\\
                          &    ivk_y     &     it_3    &              &        &       &-\epsilon_{-1}&   t_0+t_2   &              &\\
                          &    -it_3     &   -ivk_y    &     -it_3    &        &       &    t_0+t_2   &-\epsilon_{0}&   t_0-t_2    &\\
                          &              &     it_3    &     ivk_y    &        &       &              &    t_0-t_2  &-\epsilon_{+1}&\\
                          &              &             &              & \ddots &       &              &             &              &\ddots\\
\end{array}
\right)
,
\end{equation}
\end{widetext}
where the upper and lower blocks represent $s_3=\pm1$ sectors, respectively,
and $\epsilon_{n}=(-1)^n v(k_x + n q_\theta)$.
Note that $s_3=\pm1$ now represent spin polarization in $\pm y$ direction,
which is parallel to the propagating direction of the 1D interface modes.

\begin{figure}
  \begin{center}
    \includegraphics[width=1.0 \hsize]{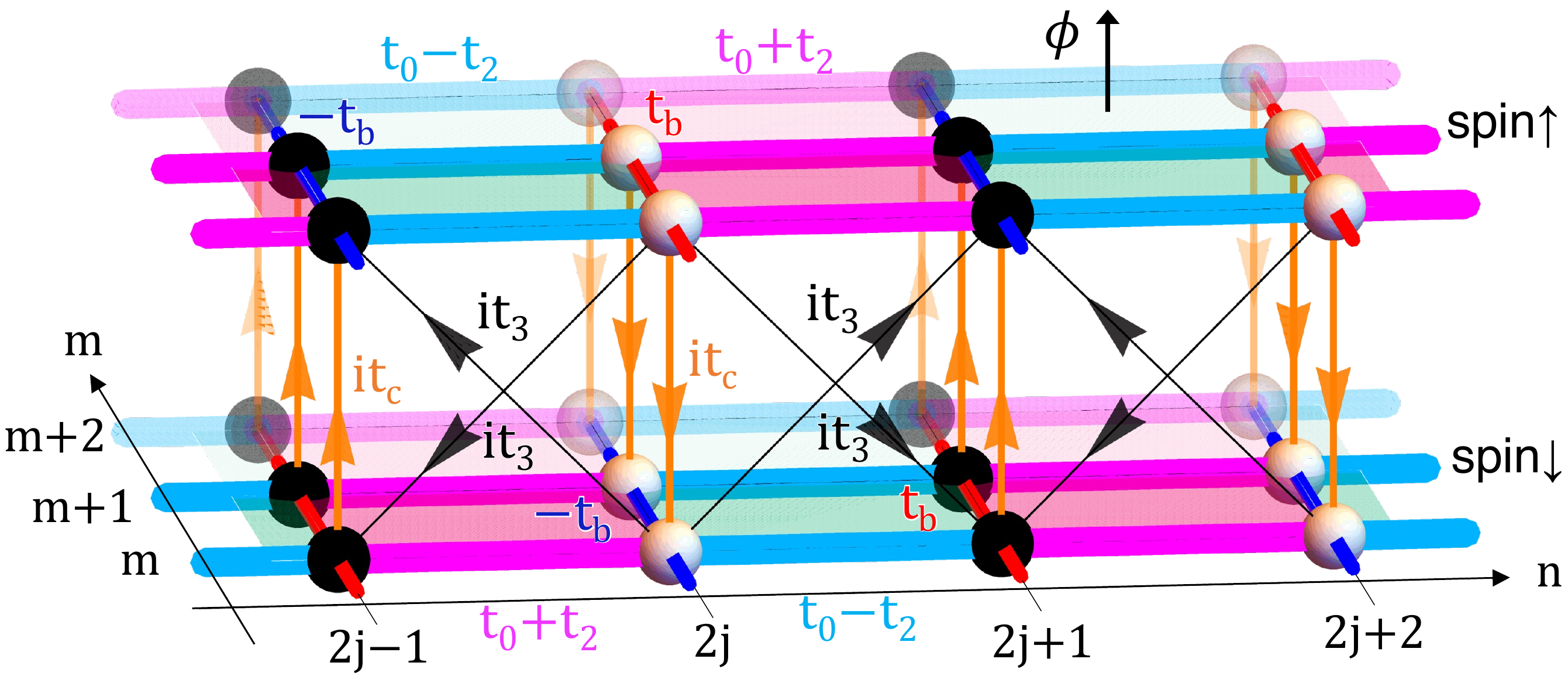}
    \caption{The square-lattice model for Eq.~\eqref{eq_bi_tb1}. The top and bottom layer represent $s_3=\pm1$, respectively. The unit cell includes a black and white site for each layer. The normal lines indicate real hopping while the arrows indicate imaginary hopping.
}
    \label{fig_landau_model}
  \end{center}
  \end{figure}

Now we show that the Hamiltonian of Eq.~\eqref{eq_Ham1_sz_pm} is equivalent to a square-lattice fermion model under a magnetic field, as illustrated in Fig.~\ref{fig_landau_model}.
Here the top and bottom layers represent spin degree of freedom $s=\uparrow, \downarrow$ (corresponds to $s_3=\pm 1$ in Eq.~\eqref{eq_Ham1_sz_pm}), respectively. The lattice points on the square grid are labeled by $(n,m)$, where the lattice spacing is 1.
The system is subjected to magnetic flux $\phi$ per $1\times1$ plaquette perpendicular to the lattice plane.
The Hamiltonian is written as
\begin{equation}\label{eq_bi_tb1}
\begin{aligned}
\mathcal{H} &=\sum_{n,m}\sum_{s=\uparrow, \downarrow} 
\Bigl[
[t_0+ s (-1)^{n} t_2] c^\dagger_{n+1,m,s} c_{n,m,s}\\
&\qquad + s (-1)^{n}\, t_b e^{2\pi in\phi} \, c^\dagger_{n,m+1,s} c_{n,m,s} \Bigr] \\
&+\sum_{n,m} \Bigl[(-1)^n \, (-it_c) \, c^\dagger_{n,m,\uparrow} c_{n,m,\downarrow} \\
&\qquad+ (-1)^n \, it_3 \, (c^\dagger_{n+1,m,\uparrow} c_{n,m,\downarrow} + c^\dagger_{n-1,m,\uparrow} c_{n,m,\downarrow})
\Bigr]\\
& + h.c.
\end{aligned}
\end{equation}
%
%
where $c^\dagger_{n,m,s}$ and $c_{n,m,s}$ 
are creation and annihilation operators of a fermion at the site $(n,m)$ with spin $s$. The variable $s$ in the equation is regarded as $\pm 1$ for $s=\uparrow, \downarrow$.
The parameters $t_0, t_2, t_3, t_b, t_c$ are all set to be real.
The hopping integrals except for $t_0$ are staggered in the $n$ direction,
The effect of the magnetic flux $\phi$ is incorporated as the Peierls phase in the hopping of $t_b$.

The Hamiltonian of Eq.~\eqref{eq_bi_tb1} is periodic along $m$, so that the eigenstates can be written as $\Psi_{n,m,s}=e^{ik_b m}\psi_{n ,s}$.
The Hamiltonian matrix for the wavefunction 
$(\cdots \psi_{-1,\uparrow},\psi_{0,\uparrow},\psi_{1,\uparrow},\cdots | 
\cdots \psi_{-1,\downarrow},\psi_{0,\downarrow},\psi_{1,\downarrow},\cdots)$
is written as
\begin{widetext} 
\begin{equation}\label{eq_Harp_Hof}
\left(
\begin{array}{ccccc|ccccc}
                   \ddots &              &             &              &        &\ddots &              &             &              &\\
                          &+\lambda_{-1} &   t_0-t_2   &              &        &       &     it_c     &    it_{3}   &              &\\
                          &    t_0-t_2   &+\lambda_{0} &    t_0+t_2   &        &       &   -it_{3}    &    -it_c    &    -it_{3}   &\\
                          &              &   t_0+t_2   & +\lambda_{+1}&        &       &              &   it_{3}    &    it_c      &\\
                          &              &             &              & \ddots &       &              &             &              &\ddots\\
                                                             \hline
                   \ddots &              &             &              &        &\ddots &              &             &              &\\
                          &     -it_c    &    it_{3}   &              &        &       &-\lambda_{-1} &   t_0+t_2   &              &\\
                          &    -it_{3}    &     it_c    &    -it_{3}   &        &       &    t_0+t_2   &-\lambda_{0} &   t_0-t_2    &\\
                          &              &    it_{3}   &     -it_c    &        &       &              &   t_0-t_2   &-\lambda_{+1} &\\
                          &              &             &              & \ddots &       &              &             &              &\ddots\\
\end{array}
\right)
\end{equation} 
\end{widetext}
where $\lambda_n=2t_b (-1)^n\cos(2 \pi \phi n+k_b)$. 
If the magnetic flux is much smaller than 1, $\lambda_n$ is approximated by $2t_b(-1)^n (k_b+2\pi n \phi)$ in the vicinity of $k_b=\pi/2$.
We immediately see that Eq.~\eqref{eq_Harp_Hof} is formally equivalent to Eq.~\eqref{eq_Ham1_sz_pm}, under a relationship shown in Table \ref{tab_map}.
Therefore, the 1D surface states in 3DTI are mapped to the Landau levels of the lattice fermion model of Eq.~\eqref{eq_bi_tb1}.


\begin{table}
\begin{tabular}{|c|c|}
\hline
{Lattice model} & {Twisted 3DTIs}  \\ 
\hline
$2t_b$  & $v$   \\
$t_c$ & -$vk_y$      \\
$2\pi \phi$ & $q_\theta$    \\  
$k_b$   & $k_x$  \\
$t_0,t_2,t_3$  & $t_0,t_2,t_3$     \\
spin ($s_i$) & spin ($s_i$) \,[After Eq.~\eqref{eq_spin_rotation}] \\
$n=2j, 2j+1$ ($\tau_i$) & top / bottom surfaces  ($\tau_i$) \\
\hline
\end{tabular}
\caption{Correspondence of parameters between the effective lattice model and the twisted 3DTIs.}
\label{tab_map}
\end{table}

Actually, the Landau levels of the lattice model 
can be analytically expressed by the low-energy approximation as follows.
First, we show that the model of Eq.~\eqref{eq_bi_tb1} in zero magnetic flux has Dirac bands in the low-energy region.
In $\phi=0$, a unit cell is given by a $2\times 1$ square in the $(n,m)$ grid. The corresponding Bloch Hamiltonian is written as
\begin{align}\label{eq_bi_Hof_Ham1}
&H(\bm{k})=
2t_0 \cos k_a (s_0 \otimes \tau_1) +
2t_2 \sin k_a (s_3 \otimes \tau_2) 
\nonumber\\
&\quad + 2t_b \cos k_b (s_3 \otimes \tau_3) +  
t_c (s_2 \otimes \tau_3)
+ 2t_3 \cos k_a (s_1 \otimes \tau_2),
\end{align}  
where $s_i$ is the Pauli matrix for the spin degree of freedom, and $\tau_i$ is that for the sublattice pseudospin of $n=2j, 2j+1$,
which corresponds to the top / bottom surfaces of the twisted 3DTIs [Eq.~\eqref{eq_Ham1_sz_pm}].
The $\bm{k}=(k_a,k_b)$ is the Bloch wave vectors for $(n,m)$ space, and
the BZ is given by $-\pi/2\leq k_a \leq \pi/2$ and $-\pi\leq k_b \leq \pi$, in accordance with the $2\times 1$ unit cell.

The energy spectrum of Eq.~\eqref{eq_bi_Hof_Ham1} has independent massive Dirac bands at $(k_a, k_b)=(\pi/2, \pm \pi/2)$.
We choose $\bm{k}_0=(\pi/2,\pi/2)$ and expand Eq.~\eqref{eq_bi_Hof_Ham1} with respect to $\bm{k}_0$.
Since $H(\bm{k})$ commutes with $s_1\tau_1$ when $t_3=0$,
we block-diagonalize the Hamiltonian into $s_1\tau_1=\pm1$ sectors by a unitary transformation. Specifically, we take the basis of $U = (\bm{u}^+_1,\bm{u}^+_2,\bm{u}^-_1,\bm{u}^-_2)$, where
\begin{align}\label{eq_base_u}
\bm{u}^\pm_1 = (z,\mp z^*,-z^*,\pm z)^T, \quad
\bm{u}^\pm_2 = (-z^*,\pm z,z,\mp z^*)^T,
\end{align}
with $z=(1+i)/(2\sqrt{2})$, and four components in each vector represent wave amplitudes of $(\tau_3,s_3) =(+,+),(+,-),(-,+),(-,-)$.
The $\bm{u}^\pm_i$ corresponds to $s_1\tau_1=\pm1$, respectively.
The transformed Hamiltonian reads 
\begin{align}\label{eq_bi_Hof_f1}
&U^\dagger H(\bm{k}) U=
\begin{pmatrix}
         h_+          &  2 i t_3k_a\sigma_2  \\
        -2 it_3k_a\sigma_2    &        h_-      \\
\end{pmatrix},
\end{align}
with
\begin{align}\label{eq_bi_Hof_m_D}
 h_{\pm}= -2 t_0 k_a\sigma_1 + 2t_b k_b\sigma_2 + M_{\pm}\sigma_3, 
\quad M_{\pm}= 2 t_2 \pm t_c,
\end{align}
where $h_{\pm}$ is for $s_1\tau_1=\pm1$, and $\sigma_i$ the Pauli matrix for the degree of freedom in each sector.
Obviously $h_{\pm}$ are 2D massive Dirac Hamiltonians.

The Landau levels in a magnetic flux $\phi$ can be obtained by replacing momentum $(k_a,k_b)$ with magnetic momentum $(\pi_a,\pi_b)$, which satisfies a commutative rule $[\pi_a,\pi_b]=-2\pi i \phi$.
By using ladder operators $a$ and $a^\dagger$ which satisfy $[a,a^\dagger]=1$,
the magnetic momentum can be expressed as
\begin{equation}
(\pi_a,\pi_b)=
\left(
-\frac{\Delta}{4t_0}
(a^\dagger+a),
-\frac{\Delta}{4it_b}
(a^\dagger-a)
\right),
\label{eq_pi_a_b}
\end{equation}
where 
$\Delta=\sqrt{16\pi t_b t_0\phi}$. 
Then the Hamiltonian $h_\pm$ becomes
\begin{equation}\label{eq_h_pm_LL}
  h_\pm =
  \begin{pmatrix}
    M_\pm & \Delta a^\dagger\\
    \Delta a & -M_\pm
  \end{pmatrix}.
\end{equation}
When $t_3=0$, the energies of the Landau levels are given by 
the eigenvalues of $h_\pm$, or
\begin{equation}\label{eq_landau_ev1}
E_{n,\pm} =
\begin{cases}
M_{\pm} & ( n= 0 ) \\
\mbox{sgn}(n) \sqrt{|n|\Delta^2 + M^2_{\pm}}& ( n=\pm1,\pm2,... )
\end{cases}.
\end{equation}
By using the mapping of Table \ref{tab_map}, we have
\begin{equation}\label{eq_twist_topo_eig}
M_\pm=2t_2 \mp vk_y,\quad \Delta = \sqrt{4 t_0 v q_\theta}.
\end{equation}

\begin{figure}
  \begin{center}
    \includegraphics[width=1.0 \hsize]{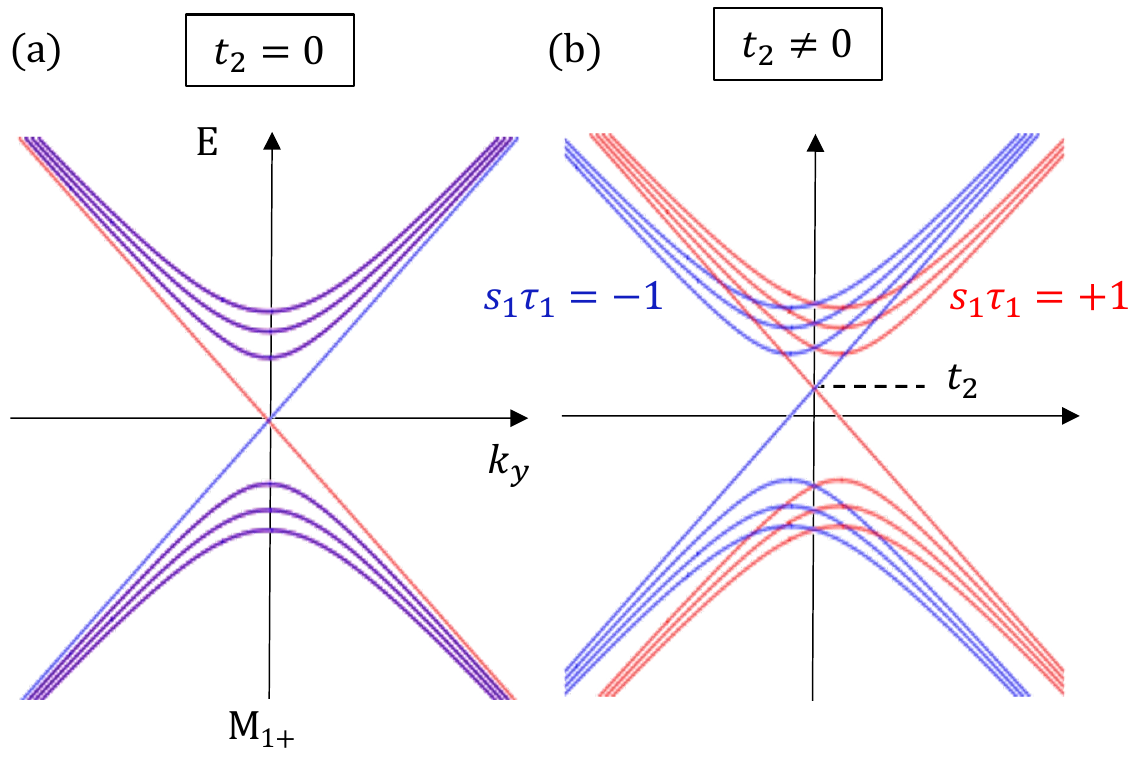}
    \caption{
    Band dispersion of the effective Landau levels of Eq.~\eqref{eq_landau_ev1} as a function of $k_y$ for (a) $t_2=0$ and (b) $t_2\neq 0$. 
    The red and blue lines represent $s_1\tau_1=\pm1$ sectors, respectively.
}
    \label{fig_landau_band}
  \end{center}
  \end{figure} 
  
This perfectly explains the band structures of the twisted 3DTIs in the small twist angle limit [Fig.~\ref{fig_band_C2xC2y}].
The flatness in $k_x$ direction corresponds to the fact that Landau level is dispersionless.
On the other hand, the band disperses in $k_y$ direction because $k_y$ corresponds to the mass parameter $M_\pm$ in the Landau level picture.
In Fig.~\ref{fig_landau_band},
we plot the band energy of Eq.~\eqref{eq_landau_ev1} as a function of $k_y$ for (a) $t_2=0$ and (b) $t_2\neq 0$, with $t_3=0$.
The red and blue lines represent the eigenenergies of $s_1\tau_1=\pm1$ sectors, respectively.
Importantly, the zeroth Landau levels $E_{0,\pm} = M_\pm$ are always linear in $k_y$ with the band velocity $\mp v$ (the original band velocity of the surface Dirac cone) regardless of other parameters.
If $t_2$ is switched on, all the energy bands in $s_1\tau_1=\pm1$ sectors horizontally shift by $\pm 2t_2/v$ because $M_\pm=2t_2\mp vk_y$.
The off-diagonal element $t_3$ hybridizes $s_1\tau_1=\pm1$ sectors, while never splits double degeneracy at $k_y=0$ because it is the Kramer's doublet in the original system.
These results are consistent with the numerical results in Fig.~\ref{fig_band_C2xC2y}.

Since $q_\theta (\propto \theta)$ corresponds to $2\pi\phi$ in Table \ref{tab_map}, a larger twist angle in the 3DTI interface corresponds to a greater magnetic field in the effective lattice model.
In the band structure of the twisted 3DTIs
[Fig.\ \ref{fig_band_C2xC2y}], indeed,
the Landau level spacing increases when $q_\theta$ is increased.
We also note that these flat levels only appear in an energy range of $|E| < 2t_0$.
This is understood by a finite depth of the Dirac cone in the effective lattice model, Eq.~\eqref{eq_bi_Hof_Ham1}, 
where the Fermi circle is closed only in $|E| < 2t_0$.
When the magnetic flux is too large, the Landau levels get broadened  as seen in Figs.\ \ref{fig_band_C2xC2y}(a)(b), because of a magnetic breakdown beyond the barrier of $2t_0$.
The condition is given by $\Delta \gtrsim 2t_0$, or, equivalently, $v q_\theta \gtrsim t_0$.
Conversely, the condition for the formation of perfect 1D channels is given by
\begin{equation}
    v q_\theta \lesssim t_0,
\label{eq_condition_perfect_1d}
\end{equation}
which agrees well with the numerical band calculation in Fig.~\ref{fig_band_C2xC2y}.

\subsection{Wavefunction of zeroth modes}\label{sec_wave}

\begin{figure}
  \begin{center}
    \includegraphics[width=1.0 \hsize]{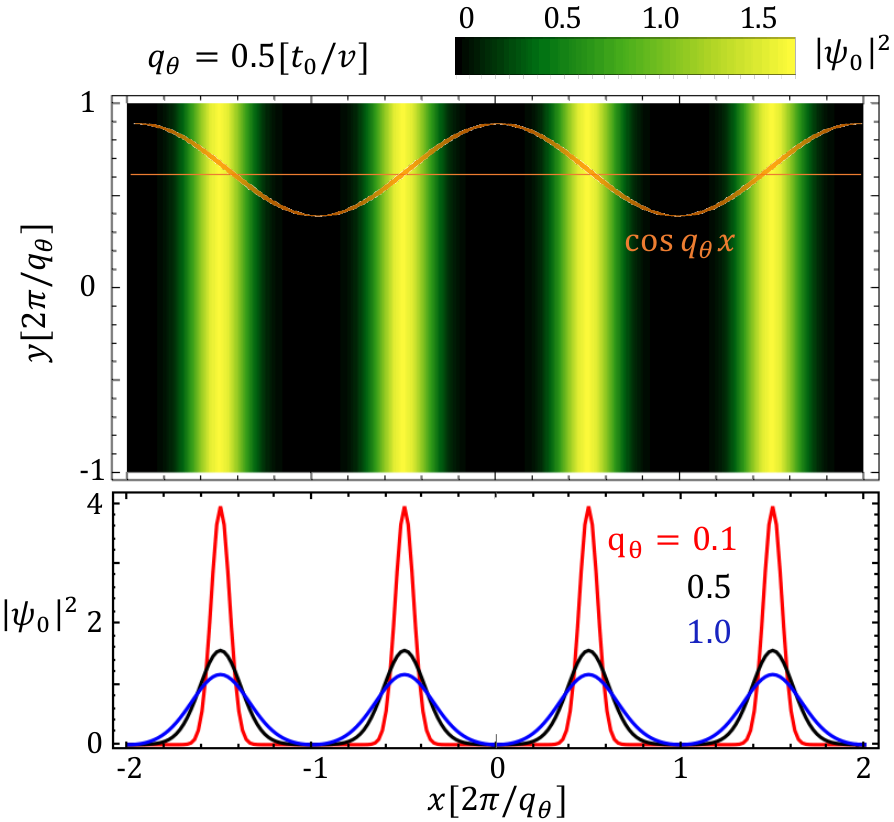}
    \caption{
    The square ampltitude of a wavefunction in the zero-th level on the twisted TI interface with $t_2=t_3=0$.
    The upper panel is a density plot on $xy$-plane for $q_\theta=0.5[t_0/v]$, where the orange curve represents $\cos q_\theta x$.
    The lower panel shows one-dimensional plot against $x$
    for $q_\theta=0.1, 0.5, 1.0 [t_0/v]$.
}
    \label{fig_1D_wave}
  \end{center}
  \end{figure} 

The wavefunctions of pseudo-LL states have highly anisotropic nature
in the real space, in accordance with the perfectly one-dimensional energy dispersion.
In Fig.~\ref{fig_1D_wave},  we plot the square amplitude of a wavefunction in the zero-th level on the twisted TI interface,
where $t_2=t_3=0$ is assumed for simplicity.
The wavefunction exhibits a stripe pattern, 
which is extended in $y$-direction while localized in $x$ direction.
The center coordinates of the stripes are corresponding to the positions where
the interlayer matrix $T_\pm(\mathbf{r}) \propto t_0 \cos \mathbf{q}_\theta\cdot\mathbf{r}$ vanishes. The spacing of the stripes is given by $l_{\rm spc} = \pi/q_\theta$, which is proportional to the moir\'{e} period.
Owing to the pseudo-Landau level description, 
the wavefunctions in $x$ direction are well approximated by $|\psi_0| \sim e^{-x^2/(2l_{\rm B}^2)}$, where 
\begin{equation}
    l_{\rm B} = \sqrt{\frac{v}{2t_0q_\theta}}
    \label{eq_l_B}
\end{equation}
is the effective magnetic length. 
The $l_{\rm B}$ is obtained by using an expression
$\pi_b =[i/(\sqrt{2}l_{\rm B})](a^\dagger -a)$ in Eq.\ \eqref{eq_pi_a_b},
and the parameter correspondence in Table~\ref{tab_map}.
The width of the stripe is estimated by the FWHM (full width at half maximum) of $e^{-x^2/(2l_{\rm B}^2)}$, which is $2\sqrt{2\log 2}\, l_{\rm B}\equiv l_w$.
The ratio of the stripe width to the stripe spacing is $l_w /l_{\rm spc} = [2\sqrt{\log 2}/\pi]\sqrt{vq_\theta/t_0}$. 
The condition that the 1D channels are spatially separated, $l_w \lesssim l_{\rm spc}$, is equivalent to Eq.~\eqref{eq_condition_perfect_1d}.

The wavefunction of the zeroth level also has a peculiar spin structure.
According to Eqs.~\eqref{eq_bi_Hof_f1} and \eqref{eq_h_pm_LL}, the eigenfunction for $n=0$ Landau levels in $s_1\tau_1=\pm 1$ sectors have the form of $\psi_{0,+} \propto (1,0,0,0)$ and $\psi_{0,-} \propto (0,0,1,0)$, 
which correspond to $\bm{u}^+_1$ and $\bm{u}^-_1$, respectively, in the original representation before the unitary transformation [Eq.~\eqref{eq_base_u}].
The $\bm{u}^\pm_1$ is fully spin-polarized in the $\pm s_2$ direction on the upper TI surface (1st and 2nd components), and in the $\mp s_2$ direction on the lower TI surface (3rd and 4th components).
Since $s_2$ corresponds to $s_1$ in the original spin axis 
[Eq.~\eqref{eq_spin_rotation}], the actual spin-polarization direction is along $\pm x$, which is parallel to the energy contour of the 1D surface band.
Since $s_1\tau_1=\pm 1$ have opposite propagating directions, we conclude that zeroth 1D interface modes in twisted 3DTIs have
counter-propagating spin currents on the upper and lower surfaces
as schematically illustrated in Fig.~\ref{fig_3Dband}.
Because of the opposite spin configurations, the left-going and right going 1D modes of $n=0$ are never hybridized by spin-independent scatters.
When the Fermi level lies in the zeroth level, therefore,
the system is equivalent to a parallel array of independent 1D channels free from the impurity scattering.

We expect that perfect 1D state is experimentally feasible in realistic materials. 
Let consider SnTe as an example of M-centered 3DTI. The parameters are given by $v=0.9\times10^6$ m/s, $t_0=0.2$ eV and lattice constant $a=0.6$
nm\cite{dziawa2012topological,hsieh2012topological,tanaka2012experimental,tanaka2013two}.
The wavenumber $q_\theta$ [Eq.~\eqref{eq_q_theta}] is given by $[2\pi/(\sqrt{3}a)]\theta$.
The condition for the 1D channel formation, Eq.~\eqref{eq_condition_perfect_1d}, then becomes
$\theta \lesssim 3^\circ$.
The width of the 1D channel is $2\sqrt{2\log 2}\, l_{\rm B} \approx (9/\sqrt{\theta^\circ})$nm [Eq.~\eqref{eq_l_B}], and the spacing is $\pi/q_\theta \approx (30/\theta^\circ)$nm.
Note that $v$ in the equations is replaced with $\hbar v$ in above estimations.

The well-separated, nanometer-scale 1D channel may exhibit Luttinger liquid behavior \cite{tomonaga1950remarks,luttinger1963exactly,haldane1981effective,voit1995one,giamarchi2003quantum,chang2003chiral}.
In the literature, arrays of parallel weakly-coupled Luttinger liquid have been extensively investigated to study non-Fermi liquids in high dimensions \cite{wen1990metallic,emery2000quantum,sondhi2001sliding,vishwanath2001two,mukhopadhyay2001sliding}, exotic quantum Hall states\cite{kane2002fractional,teo2014luttinger,tam2021nondiagonal}, topological phases\cite{neupert2014wire,iadecola2016wire} and quantum spin liquids\cite{meng2015coupled,patel2016two}. 
In our twisted 3DTI system, the coupling between 1D channels can be tuned by the twisted angle,
and it is expected to be an ideal platform to realize these novel quantum phenomena.

\section{Conclusion}
\label{sec_concl}
We have studied the electronic structure of interface states of twisted 3DTIs.
In the case of side-centered 3DTI, the surface Dirac cones are hybridized by a one-dimensional interface coupling with a single moir\'e wave number, resulting in an array of nearly-independent one-dimensional channels.
The two zeroth levels have counter-propagating spin currents free from spin-independent impurity scattering.
The one-dimensional states can be interpreted as Landau levels of the effective lattice model, where the magnetic field corresponds to the twist angle.
The coupling amplitude between the neighboring 1D channels can be controlled by the twist angle, and the system would serve as a platform for weakly-coupled parallel Luttinger liquid.

\section*{Acknowledgments}
This work was supported in part by JSPS KAKENHI Grant Number JP21H05236, JP21H05232, JP20H01840, JP20H00127, and JP20K14415 and by JST CREST Grant Number JPMJCR20T3, Japan. M.F. was supported by a JSPS Fellowship for Young Scientists. The numerical calculations were performed on XC40 at YITP in Kyoto University.

\appendix

\section{Derivation of the form of Dirac cone}
\label{sec_deriv_Dirac}
Here, we derive Eq.~\eqref{eq_surface_Dirac}, the Dirac Hamiltonian of surface mode on 3DTI in the presence of the
mirror symmetry $\mathcal{M}_{x}$ and $\mathcal{M}_{y}$.
A general $2\times 2$ Hamiltonian linear in wavenumber $(k_x,k_y)$ is written as
\begin{equation}\label{eq_general_Dirac}
H(\bm{k}) = \sum_{\nu=0,3}a_{x,\nu}s_\nu k_x + \sum_{\nu=0,3}a_{y,\nu}s_\nu k_y,
\end{equation}
where $s_1, s_2, s_3$ are Pauli matrices for spin degree of freedom and $s_0$ is a $2\times 2$ unit matrix,
and $a_{i,\nu}$ $(i=x,y)$ are real numbers.

Since the 3DTI has the time reversal symmetry, we have
\begin{equation}\label{eq_T_sym}
\mathcal{T}H(\bm{k})\mathcal{T}^{-1}=H(-\bm{k}),
\end{equation}
where $\mathcal{T}=is_2\mathcal{K}$, where $\mathcal{K}$ and the complex conjugate operator.
This immediately gives $a_{x,0}=a_{y,0}=0$.
Time reversal symmetry protects the double degeneracy of surface Dirac cone and it can emerge only at a TRIM. 

We assume that the system has both of the mirror-reflection symmetries of $\mathcal{M}_{x}$ (with respect to $yz$ plane) and $\mathcal{M}_{y}$ ($zx$ plane).
The operators are expressed as $\mathcal{M}_{x}=is_1$ and $\mathcal{M}_{y}=is_2$.
The $\mathcal{M}_{x}$ requires
\begin{equation}\label{eq_Mx_sym}
\mathcal{M}_{x}H(\bm{k})\mathcal{M}_{x}^{-1}=H(\mathcal{M}_{x}\bm{k}),
\end{equation}
resulting in $a_{x,1}=a_{y,2}=a_{y,3}=0$.
The $\mathcal{M}_{y}$ requires
\begin{equation}\label{eq_My_sym}
\mathcal{M}_{y}H(\bm{k})\mathcal{M}_{y}^{-1}=H(\mathcal{M}_{y}\bm{k}),
\end{equation}
giving $a_{x,1}=a_{x,3}=a_{y,2}=0$.
If the system has both $\mathcal{M}_{x}$ and $\mathcal{M}_{y}$, 
only $a_{x,2}$ and $a_{y,1}$ can remain non zero,
and therefore the Hamiltonian can be written in a form of
\begin{equation}\label{eq_general_Dirac_sym}
H=v_x s_2 k_x - v_y s_1 k_y,
\end{equation}
which is Eq.~\eqref{eq_surface_Dirac}.


\section{M-point 3DTI with $\mathcal{T}$ and $\mathcal{M}_x$ or $\mathcal{M}_y$}
\label{sec_T}

In the main part of the paper, we assume the 3DTI before stacking has both $\mathcal{M}_{x}$ and $\mathcal{M}_{y}$,
to show the side-centered surface Dirac cone becomes 1D channels in a twisted 3DTI.
However, most of real topological insulators actually have lower spatial symmetry. For example, the SnTe (111) surface,
which hosts the BZ side-centered Dirac cone,
has $\mathcal{M}_{x}$ symmetry while no $\mathcal{M}_{y}$
 \cite{tanaka2012experimental, PhysRevB.90.235114, PhysRevB.89.125308}.
In this appendix, we consider 
3DTIs with side-centered surface Dirac cones,
which have {\it either} of $\mathcal{M}_x$ or $\mathcal{M}_y$, to demonstrate the 1D interface states still exist.

In each case, we construct twisted 3DTIs in the same manner as in Fig.~\ref{fig_schem}.
We flip one slab and stack it on the top of the other,
so that two slabs share the same mirror plane
($yz$-plane for $\mathcal{M}_{x}$ and 
$zx$-plane for $\mathcal{M}_{y}$).
Then we twist the top and bottom slabs with respect the $z$-axis by $\pm\theta/2$, respectively.
The entire system has $\mathcal{C}_{2x}(\mathcal{C}_{2y})$ if the slab has $\mathcal{M}_x(\mathcal{M}_y)$.

\subsection{$\mathcal{T}$ and $\mathcal{M}_x$}

We consider the case where the 3DTI slab has $\mathcal{T}$ and $\mathcal{M}_{x}$. 
This applies to SnTe (111) surface.
From the symmetry constraints $\mathcal{T}$ [Eq.~\eqref{eq_T_sym}] and $\mathcal{M}_{x}$ [Eq.~\eqref{eq_Mx_sym}], the surface Dirac Hamiltonian $H_{u/l}$ is written in the lowest order in $k$ as 
\begin{equation}\label{eq_general_Dirac_Mx}
H_{u/l}=\pm [(a_{x2} s_2 + a_{x3} s_3) k_x + a_{y1} s_1 k_y].
\end{equation}
where $\pm$ correspond to $u$ and $l$, respectively.
If we rotates the spin axis on $s_2$-$s_3$ plane, we can make the spin texture lie on a $s_1$-$s_2$ plane and the Dirac Hamiltonian become the same form of Eq.~\eqref{eq_general_Dirac_sym}.

Now the interface-coupling Hamiltonian takes the form of $H^{(M_1)}_{\rm int} = T_+ e^{i\bm{q}_\theta \cdot \bm{r}} + T_- e^{-i\bm{q}_\theta \cdot \bm{r}}$ as argued in Sec.~\ref{sec_eff_3DTI}.
By requiring that the twist 3DTI model respects $\mathcal{T}$ and $\mathcal{C}_{2x}$,
the interlayer Hamiltonian Eq.~\eqref{eq_general_T} is reduced to
\begin{align}\label{eq_M_TCx_app}
T_\pm = t_0 s_0 + (\pm t_2 + i t_2^{\prime}) s_2 + (\pm t_3^{\prime} + i t_3) s_3,  
\end{align} 
where $t_2^\prime$ and $t_3^{\prime}$ are real numbers.
Note that the general form of the interface Hamiltonian does not change in the spin rotation on $s_2$-$s_3$ plane argued above.

\begin{figure*}[t]
 \begin{center}
  \includegraphics[width=1 \hsize]{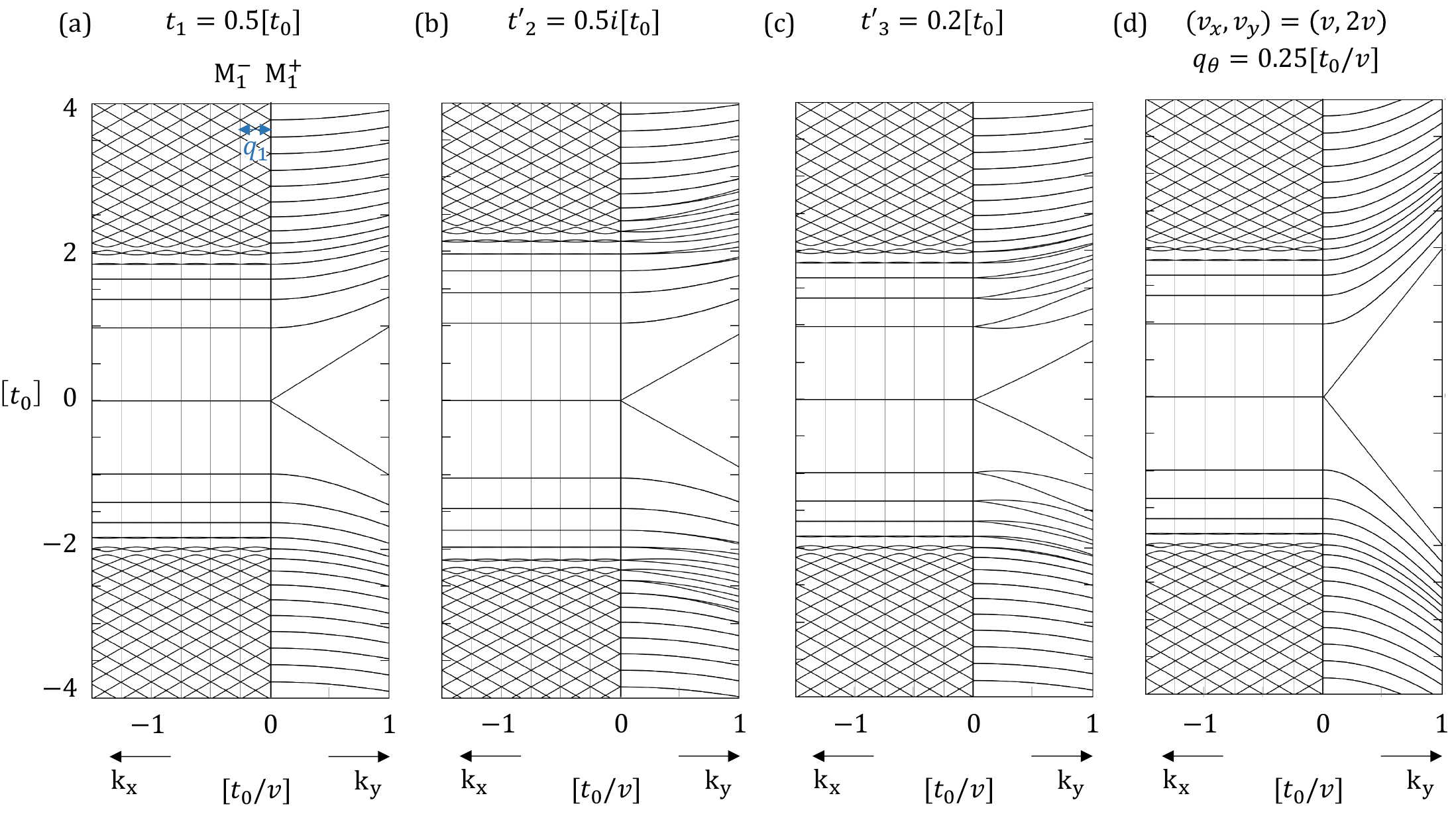}
  \caption{
Plots similar to Fig.~\ref{fig_band_C2xC2y}(c) for (a)$t_1/t_0=0.5$, (b)$t_2^\prime/t_0=0.5$ and (c)$t_3^\prime/t_0=0.2$. (d) The band structure for the anisotropic Dirac cone at $M_1$ with $(v_x,v_y)=(v,2v)$ and $q_\theta = 0.25(t_0/v)$.
 }
  \label{fig_band_app}
 \end{center}
\end{figure*}

Here, we investigate the effect of $t_2^\prime$ and $t_3^\prime$ which are not discussed in Sec. \ref{sec_band}.
In the following, we assume an isotropic surface Dirac cone $v_x=v_y (\equiv v)$ for simplicity.
By diagonalizing Eq.~\eqref{eq_M_Ham}, we numerically calculate the enregy bands of the twisted 3DTIs with the surface Dirac cone at $M_1$. 
The band structures for nonzero $t_2^\prime$ and $t_3^\prime$ are plotted in Fig.~\ref{fig_band_app}(b) and (c), respectively, with 
$q_\theta = 0.25(t_0/v)$.
We observe that $t_2^\prime$ hardly changes the entire band structure, while $t_3^\prime$ splits the double degeneracy in $k_y\neq0$.
If we consider the pseudo-Landau mapping, the corresponding Hamiltonian to Eq.~\eqref{eq_bi_Hof_f1} is given as
\begin{align}
\begin{pmatrix}
                           h_+                        &  2 i t_3k_a\sigma_2 - 2 i t_3^\prime\sigma_2  \\
        -2 it_3k_a\sigma_2 + 2 i t_3^\prime\sigma_2    &                    h_-                       \\
\end{pmatrix},
\end{align}
with
\begin{equation}
  \begin{split}
&h_{\pm}= -2 t_0 k_a\sigma_1 + 2t_b k_b\sigma_2 + M_{\pm}\sigma_3,  \\
&\qquad M_{\pm}= 2 t_2 - 2 t_2^\prime k_a \pm t_c.
  \end{split}
\end{equation}
The $t_2^\prime$ is coupled with $k_a$ so that the Dirac cone of $h_{\pm}$ is unchanged in the vicinity of $k_a=0$.
The $t_3^\prime$ hybridizes $s_1 \tau_1=\pm1$ sectors just as $t_3$ does. These results are consistent with the numerical calculations in Fig.~\ref{fig_band_app}.
Therefore, the 1D flat bands are expected to be observed in 
the twisted stack of 3DTI having $\mathcal{M}_x$ only, such as
SnTe $(111)$ surface.

\subsection{$\mathcal{T}$ and $\mathcal{M}_y$}

We assume that the bulk of nontwisted 3DTI has $\mathcal{T}$ and $\mathcal{M}_{y}$, resulting in $\mathcal{T}$ and $\mathcal{C}_{2y}$ for the twisted 3DTIs.
From the symmetry constraints $\mathcal{T}$ [Eq.~\eqref{eq_T_sym}] and $\mathcal{M}_{y}$ [Eq.~\eqref{eq_My_sym}], the surface Dirac Hamiltonian $H_{u/l}$ is written in the lowest order in $k$ as 
\begin{equation}\label{eq_general_Dirac_My}
H_{u/l}=\pm a_{x2} s_2 k_x \pm (a_{y1} s_1 + a_{y3} s_3) k_y.
\end{equation}
where $\pm$ are for $u$ and $l$.
If we rotates the spin axis on $s_1$-$s_3$ plane, the spin texture lies on a plane and the Dirac Hamiltonian become the same form of Eq.~\eqref{eq_general_Dirac_sym}.

The interface-coupling Hamiltonian takes the form of $H^{(M_1)}_{\rm int} = T_+ e^{i\bm{q}_\theta \cdot \bm{r}} + T_- e^{-i\bm{q}_\theta \cdot \bm{r}}$ as argued in Sec.~\ref{sec_eff_3DTI}.
By requiring that the twist 3DTI model respects $\mathcal{T}$ and $\mathcal{C}_{2y}$,
the interlayer Hamiltonian Eq.~\eqref{eq_general_T} is reduced to
\begin{align}\label{eq_M_TCy_app}
T_\pm = t_0 s_0 + i t_1 s_1 \pm t_2 s_2 + i t_3 s_3,  
\end{align} 
where $t_1$ is real number.
Again, the general form of the interface Hamiltonian does not change in the spin rotation on $s_1$-$s_3$ plane.

We investigate the effect of $t_1$ which is not discussed in Sec. \ref{sec_band}.
In the following, we assume an isotropic surface Dirac cone $v_x=v_y (\equiv v)$ for simplicity.
The band structure of the twisted 3DTIs with nonzero $t_1$ in Fig.~\ref{fig_band_app}(a) with $q_\theta = 0.25(t_0/v)$. 
The energy spectrum in Fig.~\ref{fig_band_C2xC2y}(c) is not modified by the $t_1$.
It can be understood by considering the pseudo-Landau mapping, where $t_1$ is included in Eq.~\eqref{eq_bi_Hof_m_D} as
\begin{equation}\label{eq_bi_Hof_m_D_app}
  \begin{split}
 &h_{\pm}= -2 t_0 k_a\sigma_1 + 2(t_b k_b \pm t_1 )\sigma_2 + M_{\pm}\sigma_3, \\
&\qquad \qquad \quad M_{\pm}= 2 t_2 \pm t_c,
  \end{split}
\end{equation}
We see $t_1$ just shifts the origin of $k_b$ (corresponding to $k_x$), and hence the 1D bands remain intact.

\subsection{Anisotropic surface Dirac cones}
\label{sec_anisotropic}

The surface Dirac cone at $M$ point (BZ side center) is generally anisotropic, i.e., $v_x \neq v_y$. This is because 
the rotation symmetry $C_n$ $(n \geqq 3)$ is absent
in a set of spatial transformations to keep the $M$ point invariant.
Here, we calculate the moir\'e band structure for the anisotropic surface Dirac cone.
For simplicity we assume the interface Hamiltonian as 
$H^{(M_1)}_{\rm int}  = t_0 s_0 \cos{(q_\theta x)}$.
Figure~\ref{fig_band_app}(d) shows the band structure for 
$(v_x,v_y)=(v,2v)$ and $q_\theta = 0.25(t_0/v)$.
Compared to Fig.~\ref{fig_band_C2xC2y}(c), the energy spectrum along $k_x$ axis is completely same while the band velocity along $k_y$ axis becomes twice.
Therefore the formation of 1D interface states is not affected by an anisotropy of the surface Dirac cone.

\bibliography{twisted_topo.bib}

\end{document}